\documentclass[12pt]{iopart}

\usepackage{iopams}  
\usepackage{cite}
\usepackage{xspace}
\usepackage{graphicx}
\newcommand{\adms}{all-dielectric metamaterials\xspace}
\newcommand{\Adms}{All-dielectric metamaterials\xspace}
\newcommand{\fer}{$f_{er}$\xspace}
\newcommand{\fmr}{$f_{mr}$\xspace}
\newcommand{\aer}{$A_{er}$\xspace}
\newcommand{\amr}{$A_{mr}$\xspace}

\newcommand{\neff}{n_{eff}\xspace}

\newcommand{\lp}{$l_p$\xspace}
\newcommand{\tio}{$\mathrm{TiO}_2$}
\newcommand{\sto}{$\mathrm{SrTiO}_3$}
\newcommand{\ntio}{$N_{TiO_{2}}$}

\newcommand{\tgd}{$\tan\delta$}

\begin{document}

\title[Negative index and mode coupling in \adms at THz frequencies]{Negative index and mode coupling in \adms at terahertz frequencies}

\author{\' Eric Akmansoy \& Simon Marcellin}

\address{Institut d'\'Electronique Fondamentale, Univ. Paris-Sud, Universit\'e Paris-Saclay, Orsay, F-91405 ; UMR8622, CNRS, Orsay, F 91405.}
\ead{eric.akmansoy@u-psud.fr}

\begin{abstract}
We report on the role of the coupling of the modes of Mie resonances in \adms to ensure negative effective index at terahertz frequencies. We study this role according to the lattice period and according to the frequency overlapping of the modes of resonance. We show that negative effective refractive index requires sufficiently strong mode coupling and that for even more strong mode coupling, the first two modes of Mie resonances are degenerated; the effective refractive index is then undeterminded. We also show that adjusting the mode coupling leads to near-zero effective index, or even null effective index. Further, we compare the mode coupling effect with  hybridization in metamaterials. 
\end{abstract}

%
%
%
%
%

\section{Introduction}

\Adms (ADM) are the promising ``inflection'' of metamaterials to go beyond their limits. 
ADMs are an alternative to  metallic metamaterials. The advantages of ADMs come from their low losses and their simple geometry: they do not suffer from ohmic losses and thus benefit from low energy dissipation. From the microwave, their quality factor is greater than that of metallic metamaterials~\cite{prl100_popa} and it is consequently at terahertz and optical frequencies. 
ADMs are partly inspired by the work of Richtmyer who developed the theory of dielectric resonators, which is based on the fact that ``the dielectric has the effect of causing the electromagnetic field [\dots] to be confined to the cylinder itself and the immediately surrounding region of space.''~\cite{jap10_richtmyer}. 
Taking the matter further, O'Brien and Pendry opened the way for ADMs by considering the periodic lattice of  high permittivity resonators (HPR), thus  evidencing artificial magnetism in the microwave~\cite{jpcm15_pendry}. 
ADMs rely on  the first two modes of Mie resonances of HPR.  The first mode results in resonant effective permeability which can have negative values, while the second one results in resonant effective permittivity which can also have negative values. When the two are simultaneously negative, the ADM is called ``double negative'' (DNG) and its effective refractive index is then negative\,~\cite{pt6_pendry_smith, mt9_smith, prl84_smith, ap51_ziolkowski, mtt53_ziolkowski, apl90_smith}. 
The unit cell of ADM thus comprises two subwavelength building blocks of simple geometry~\cite{prb77_mosallaei, mt12_lippens}. By analogy with chemical molecules, the unit cell is generally called \emph{meta-dimer}~\cite{np3_giessen}, and the two building blocks are called \emph{meta-atoms}. As the two are different, the unit cell is a hetero-dimer. 

\begin{figure}[!htb]
\begin{center}
\includegraphics[width = 0.9\linewidth]{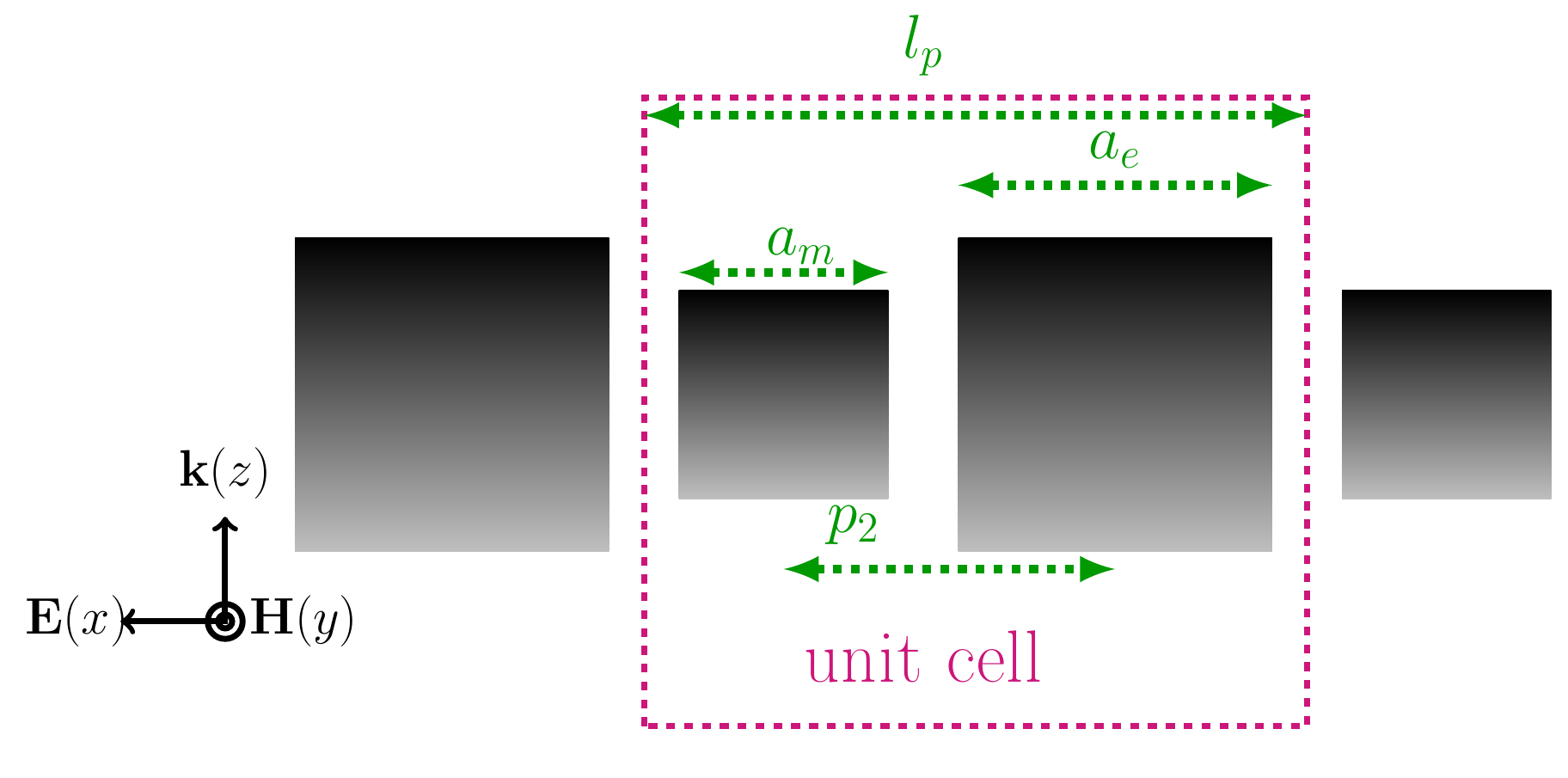}
\caption{Schematic layout of the ADM. The ADM consists of one infinite layer along the $x$ direction, and is made up of two interleaved sets of high permittivity square cross-section dielectric cylinders which resonate in the first two modes of Mie resonances: the small set resonates in the \textit{magnetic} mode and the second one in the \textit{electric} mode. The equidistant cylinders are infinite along the $y$ direction and their side lengths are $a_m= 60\,\mu m$ and $a_e = 90\,\mu m$, respectively. Their relative permittivity is $\epsilon_r = 94$. The unit cell actually consists of two subwavelength distinct building blocks. The lattice period is $l_p = 260\,\mu m$ and $p_2$ is half the lattice period.  The incident wave is transverse electric\,(TE). }
\label{unitcell}
\end{center}
\end{figure}

The large applications of ADMs (for a review, see ref.\,\cite{nn1_jacob}) include perfect reflectors~\cite{acsp2_valentine}, perfect absorbers~\cite{apl103_lippens, oe24_zhou}, zero-index metamaterials~\cite{np7_valentine}, optical magnetic mirror~\cite{o1_brenner}, Fano resonances~\cite{nm9_giessen, prb82_lepetit}. 
ADMs have been demonstrated from microwave to optical domain. Artificial magnetism~\cite{prl101_zhao, prl100_popa, el44_lepetit, m3_vendik} and negative effective refractive index\cite{prb76_kim,  apl95_lepetit} have been, theoretically and experimentally, evidenced in GHz range. Even though artificial magnetism provided by ADMs has been experimentally demonstrated in the THz range~\cite{prb79_nemec} and in the infrared range~\cite{prb79_wheeler, prl108_ginn, prl99_schuller, am24_meseguer},  DNG refractive index has not been yet demonstrated, which impede ADMs to be the equivalent of their metallic counterpart. Besides, terahertz (THz) radiation is widely defined as electromagnetic radiation in the frequency range 0.3-3 THz. Since it permits to obtain physical data which are not accessible using X-ray or infrared radiation, it finds many applications in imaging, security, quality inspection, chemical sensing, astronomy, etc. On their part, metamaterials have evolved towards the implementation of photonic components~\cite{nm11_zheludev}. HPRs are well suited for metamaterials applications in the low THz frequency range~\cite{jstqe99_gaufillet}.

In the following, we report on the mode coupling effect which plays a dominant role in the electromagnetic properties of metamaterials~\cite{acie49_giessen, njp14_lippens, pier132_lippens, josab31_capolino}, notably, in the achievement of negative effective index. 
Magnetic and electric mode coupling effects in ADMs have been separately studied from each other in the microwave by Zhang \textit{et al}\,\cite{njp14_lippens}. Herein, we report on the magneto-electric mode coupling effects going up the terahertz domain, and we show that negative effective refractive index requires sufficiently strong mode coupling~\cite{aemmo2015_marcellin}.
We also show that adjusting the mode coupling allows to attain near-zero values of the refractive index, or even null effective index. Moreover, we highlight that the strongest values of the mode coupling lead  to frequency mode degeneracy, for which the refractive index is undetermined.

\section{Simulations}
We considered a 2D ADM whose unit cell consists of two distinct building blocks, a \textit{magnetic} block and an \textit{electric} one, the former resonating in the first mode of Mie resonances and the latter resonating in the second mode\,~\cite{prl100_popa, mt12_lippens, apl95_lepetit, jap109_lepetit}. The first mode is thus referred to as the \textit{magnetic} mode and the second one to as the \textit{electric} mode. 
The ADM consists of one infinite layer made up of two sets of high permittivity square cross-section dielectric cylinders which are perpendicular to the incident wave vector (Fig.\,\ref{unitcell}). The two sets of HPRs are actually interleaved. The incident polarization is Transverse Electric (TE), i.e., the electric field is perpendicular to the axis of the cylinders. We studied both \textit{spatial} mode coupling and \textit{frequency} mode coupling. 
The ADM has been numerically simulated by the means of the finite elements method software HFSS\texttrademark\, which yields the $S$-parameters. The side lengths of the resonators are initially $a_m= 60\,\mu m$ for the \textit{magnetic} mode and $a_e = 90\,\mu m$ for the \textit{electric} one, the lattice period being $l_p = 260\,\mu m$. The HPRs are equidistant and therefore, the distance between two of them is half the lattice period $p_2=l_p/2 = 130\,\mu m$. The relative permittivity of the dielectric is $\epsilon_r = 94$ (Titanium dioxyde - \tio) and the loss tangent increases between \tgd = 0.009 and 0.015 in the considered frequency range\,\cite{jjap47_matsumoto, mtt53_berdel}. Thus, we are dealing with a high refractive index bulk material (\ntio$\simeq$ 10)

\begin{figure}[!htb]
\begin{center}
\includegraphics[width = 0.9\linewidth, trim = 0cm 1cm 0cm 1cm, clip = true]{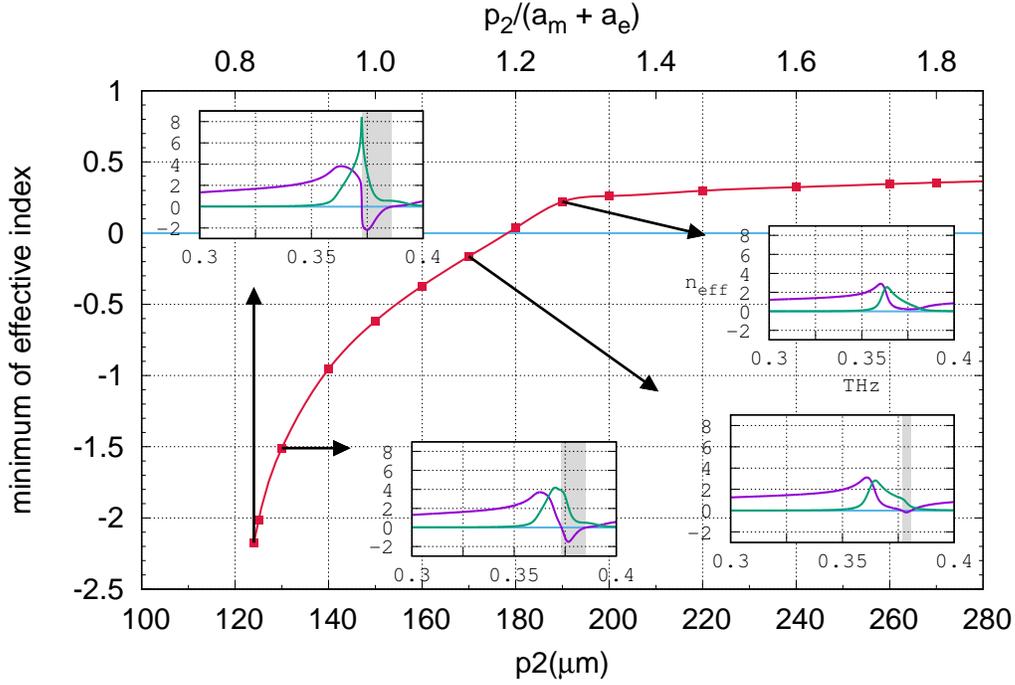}\\
\caption{\textit{Spatial} mode coupling: minimal value of the effective refractive  index $n_{eff}$ in function of the distance $p_2$ between two resonators. The side lengths of the resonators are $a_m= 60\,\mu m$ and $a_e= 90\,\mu m$, respectively.  Insert: real and imaginary part of the effective index $n_{eff}$; the shaded area denotes the bandwidth of negative effective index.}
\label{n_min_spatial}
\end{center}
\end{figure}

\begin{figure}[!htb]
\begin{center}
\includegraphics[width = 0.9\linewidth, trim = 0cm 1cm 0cm 1cm, clip = true]{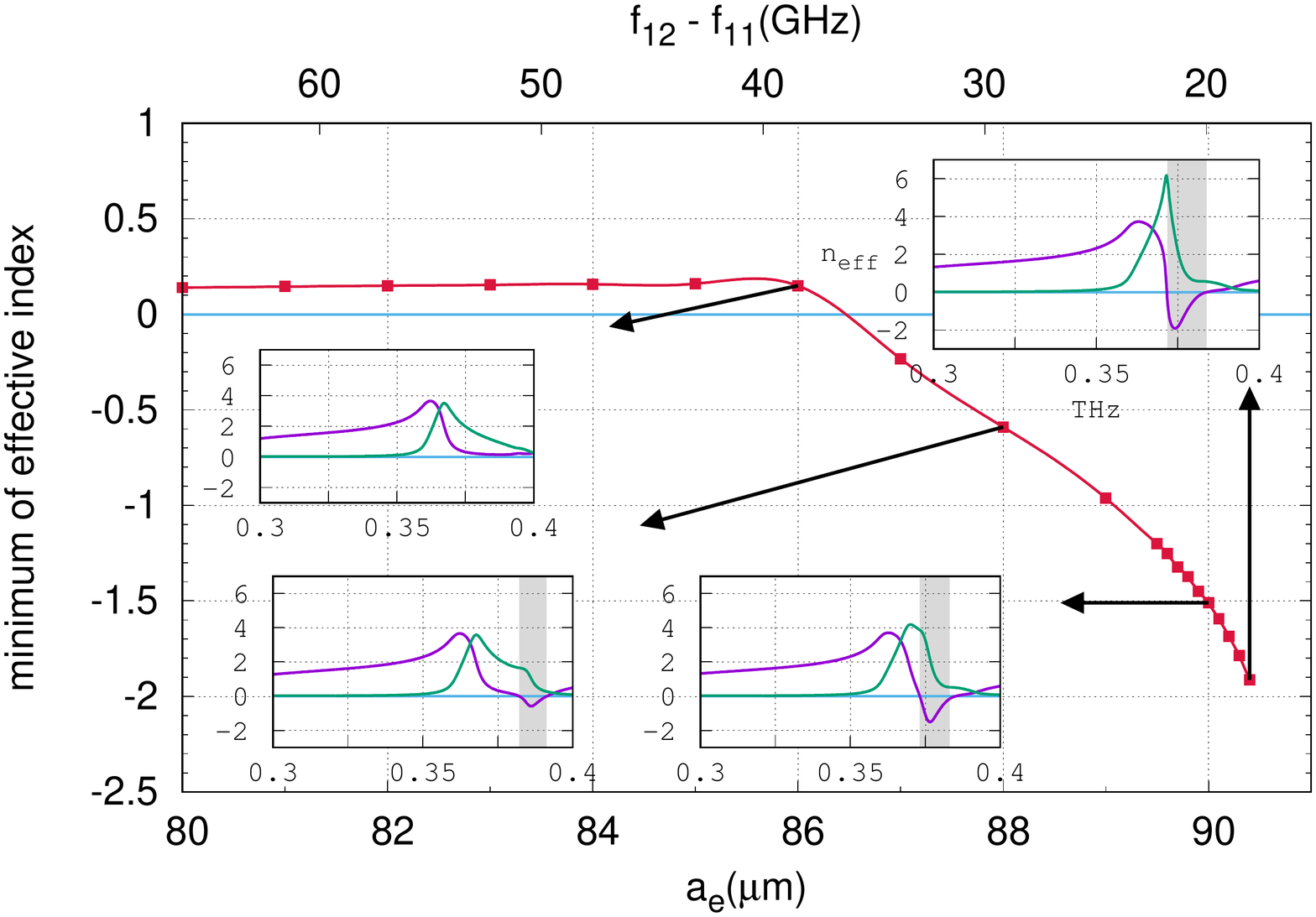}
\caption{\textit{Frequency} mode coupling: minimal value of the effective refractive  index $n_{eff}$ in function of the side length $a_e$ of the \textit{electric} resonator, namely, the frequency of the \textit{electric} mode is varying. Insert: real and imaginary part of the effective index $n_{eff}$; the shaded area denotes the bandwidth of negative effective index. $f_{11}$ and $f_{12}$ are the resonances frequencies of the individual resonators (see the text). }
 \label{n_min_freq}
\end{center}
\end{figure}

\section{Results and discussion}
\subsection{Negative index and mode coupling}
We studied the mode coupling between the first two modes of Mie resonances depending on the lattice period \lp (\textit{spatial} mode coupling), and then depending on the frequency overlapping of the two modes (\textit{frequency} mode coupling). The results of the simulation, namely the minimum $n_{eff_{min}}\xspace$ of the effective index $\neff$ in function the lattice period and in function of the frequency spacing between the two modes, are reported in Fig.\,\ref{n_min_spatial} and \ref{n_min_freq}, respectively.  They show that the mode coupling should be strong enough to ensure negative effective  index. The minima of the effective index $\neff$ are $n_{eff_{min}}\xspace$ = -2.2 and -1.9, respectively.
In the one hand, increasing the mode coupling is obtained by decreasing the lattice period $l_p$.  On the other hand, it is obtained by decreasing the frequency of the second mode of Mie resonances, which derives from the increasing of the side length $a_e$ of the \textit{electric} resonator. 
%
When the mode coupling is sufficient, the bandwidth of the negative effective index $n_{eff}$ increases with it (see insets in Fig.\,\ref{n_min_spatial} \& \ref{n_min_freq}). The frequency range of negative effective index is given by the relation~\cite{motl41_depine}
\begin{eqnarray}
\epsilon ' (\omega) \cdot \mu '' (\omega) + \epsilon '' (\omega) \cdot \mu ' (\omega) \leq 0,
\label{nneg}
\end{eqnarray}
where $'$ and $''$ respectively denote the real and imaginary parts of the permeability $\mu$ and the permittivity $\epsilon$.

\begin{figure}[!htb]
\begin{center}
\includegraphics[width = \linewidth, trim = 1cm 12cm 1cm 0cm, clip = true]{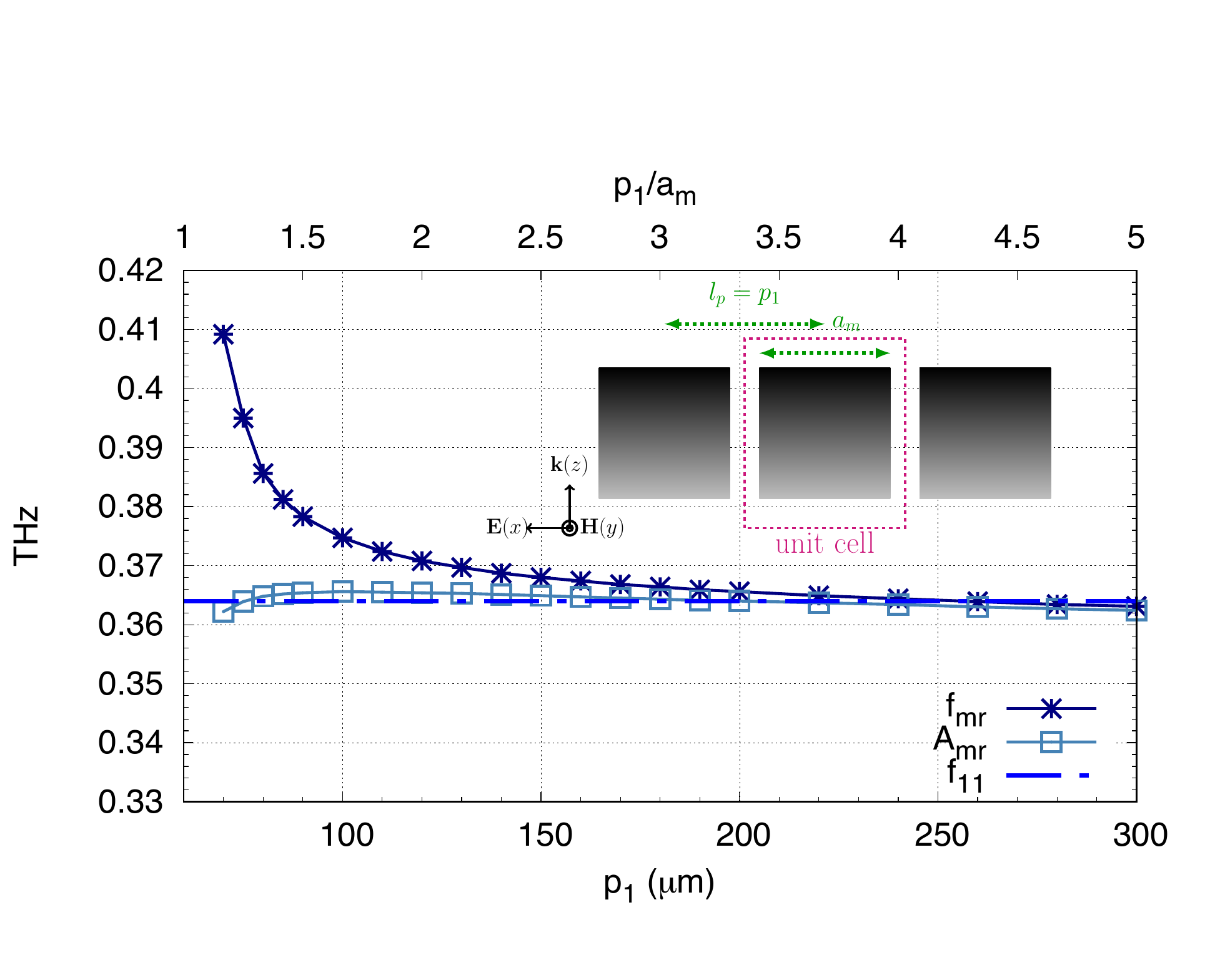}\\
\caption{\textit{Magnetic} mode coupling:\,frequency (\fmr) of the first mode of Mie resonances and of the maximum of absorption (\amr) in function of the distance $p_1$ between two resonators. The unit cell only comprises the \textit{magnetic} block. The side length of the resonator is $a_m= 60\,\mu m$. The dashed line denotes the frequency of resonance ($f_{11}$) of the mode of resonances of the individual resonator.}
\label{couplage_monomode_mag}
\end{center}
\end{figure} 
\subsection{Monomode coupling}
\begin{figure}[!htb]
\begin{center}
\includegraphics[width = \linewidth, trim = 1cm 12cm 1cm 0cm, clip = true]{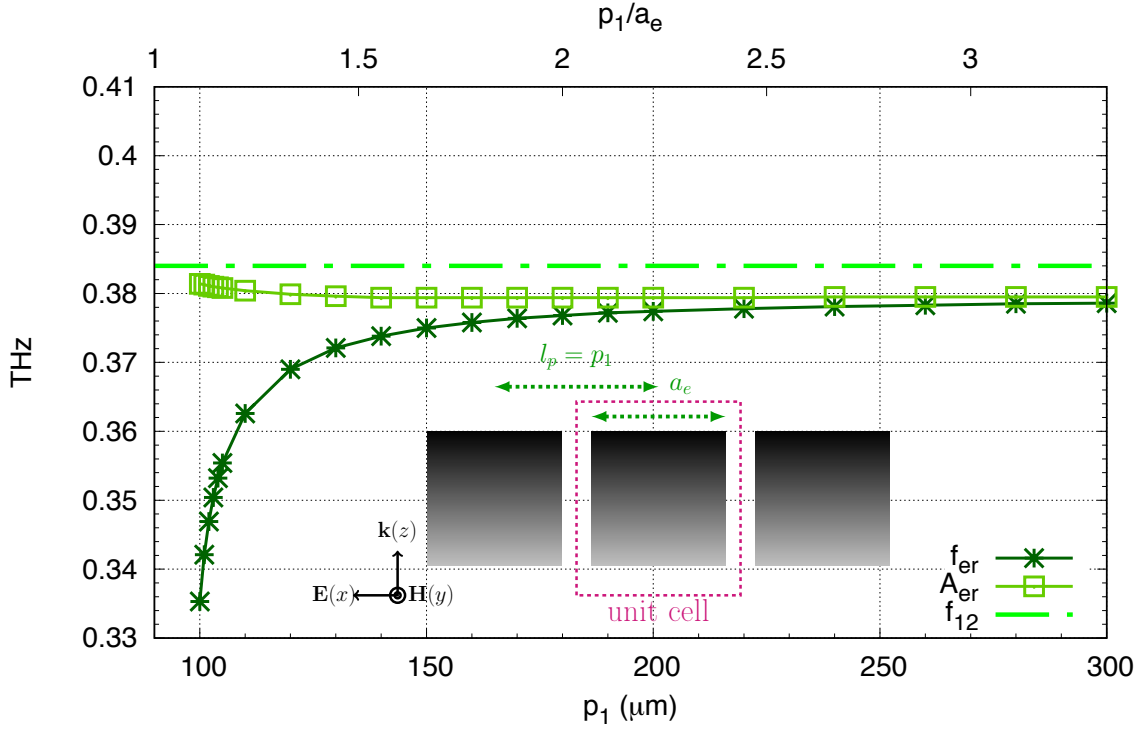}
\caption{\textit{Electric} mode coupling:\,frequency (\fer) of the second mode of Mie resonances and of the maximum of absorption (\aer) in function of the distance $p_1$ between two resonators. The unit cell only comprises the \textit{electric} block.  The side length of the resonator is $a_e= 90\,\mu m$. The dashed line denotes the frequency of resonance ($f_{12}$) of the mode of resonances of the individual resonator.}
\label{couplage_monomode_elec}
\end{center}
\end{figure}

To carry out our study, we firstly studied  monomode coupling, that is, the mode coupling due to only one mode, which arises inside a layer whose unit cell only consists of one building block, the \textit{magnetic} one or the \textit{electric} one. We studied both cases. Consequently, we had to only consider the \textit{spatial} mode coupling and we only varied the lattice period $l_p$ which is equal to the distance $p_1$ between two resonators, $l_p= p_1$. Operating in the same frequency range, the side length of the \textit{magnetic} resonator is $a_m = 60\,\mu m$, while that of the \textit{electric} resonator is $a_e = 90\,\mu m$. 
The results of the simulation are reported in Fig.\,\ref{couplage_monomode_mag} and \ref{couplage_monomode_elec}, for both cases, and they show that the two modes differently behave. Their respective frequencies (\fmr, \fer) are given by the minima of the $S_{12}$\,parameter\,~\cite{njp14_lippens}. The frequency \fmr of the \textit{magnetic} mode increases with the lattice period $l_p$, whereas the frequency \fer of the \textit{electric} one decreases. These results are consistent with that of reference\,\cite{njp14_lippens}.
Both maxima of the absorption $A_{mr}$ and $A_{er}$ ($A = 1 - \left| S_{12}\right|^2 - \left| S_{11}\right|^2$) and the frequency of the resonance modes of the individual resonator are also reported. This latter provides a series of resonances whose frequency is  determined by Cohn's model\,\cite{mtt16_cohn, guillon_book, mtt14_sethares}
%
\begin{eqnarray}\label{eq2}
f_{m\,n} = \frac{c}{2\sqrt{\epsilon_r }}\sqrt{\left(\frac{m}{a}\right )^2 + \left(\frac{n}{b}\right )^2},
\end{eqnarray}
where $\epsilon_r $ is the relative permittivity of the resonator, $m$ and $n$ are integers, $a$ and $b$ are the side lengths of the resonator and $c$ is the velocity of light and its accuracy is about 5\%. Equation\,\ref{eq2} was used to design all the reported structures. For square cross section cylinder, $a =b$, and the frequencies of the first two modes of the individual resonator respectively correspond to $m = n = 1$ and $m = 1$ and $n = 2$. To exhibit negative refractive index,  the involved resonance modes are the first mode of the \textit{magnetic} resonator and the second mode of the \textit{electric} resonator. Their frequencies are respectively $f_{11} = 0.364$\,THz ($a = a_m = 60\,\mu m$) and $f_{12} = 0.384$\,THz ($a = a_e = 90\,\mu m$)~\cite{apl95_lepetit}.  
These are obviously constant, while the maxima of the absorption are practically constant. It can be noticed that, as the mode coupling increases, the distance between the resonance frequencies (\fmr, \fer) of the resonator inside the layer and that ($f_{11}$, $f_{12}$)  of the individual resonator respectively increases (cf.\,Fig.\,\ref{couplage_monomode_mag} and \ref{couplage_monomode_elec}, respectively), thus evidencing the mode coupling. 

\subsection{Bimode coupling}
\begin{figure}[!htb]
\begin{center}
\includegraphics[trim = 0cm 1cm 0cm 0cm, clip = true, width = \linewidth] {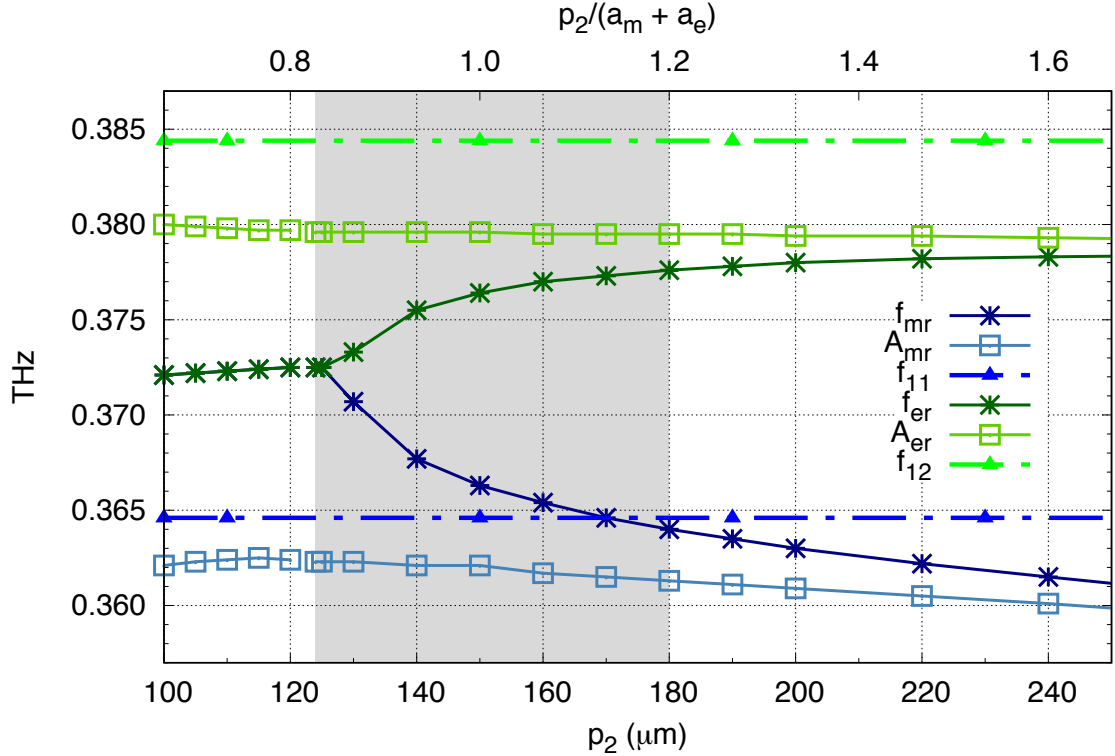}
\caption{\textit{Spatial} mode coupling:\,frequency of the first two modes of Mie resonances in function of the distance $p_2$ between two resonators which is half the lattice period $l_p$.
Blue and green colors denote the \textit{magnetic} and the \textit{electric} modes, respectively. Square dots and crossed dots denote the maxima of absorption (\amr, \aer) and the minimum of the {$S_{12}$}\,parameter (\fmr, \fer), respectively. The shaded area corresponds to negative value of the effective index $\neff$. The side lengths of both resonators are $a_m= 60\,\mu m$ et $a_e = 90\,\mu m$, respectively. The dashed lines denote the frequencies of resonances ($f_{11}$, $f_{12}$) of the modes of the individual resonator. 
}
\label{diapason_pas}
\end{center}
\end{figure}

\begin{figure} [!htb]
\begin{center}
\includegraphics[trim = 0cm 1cm 0cm 0cm, clip = true, width = \linewidth] {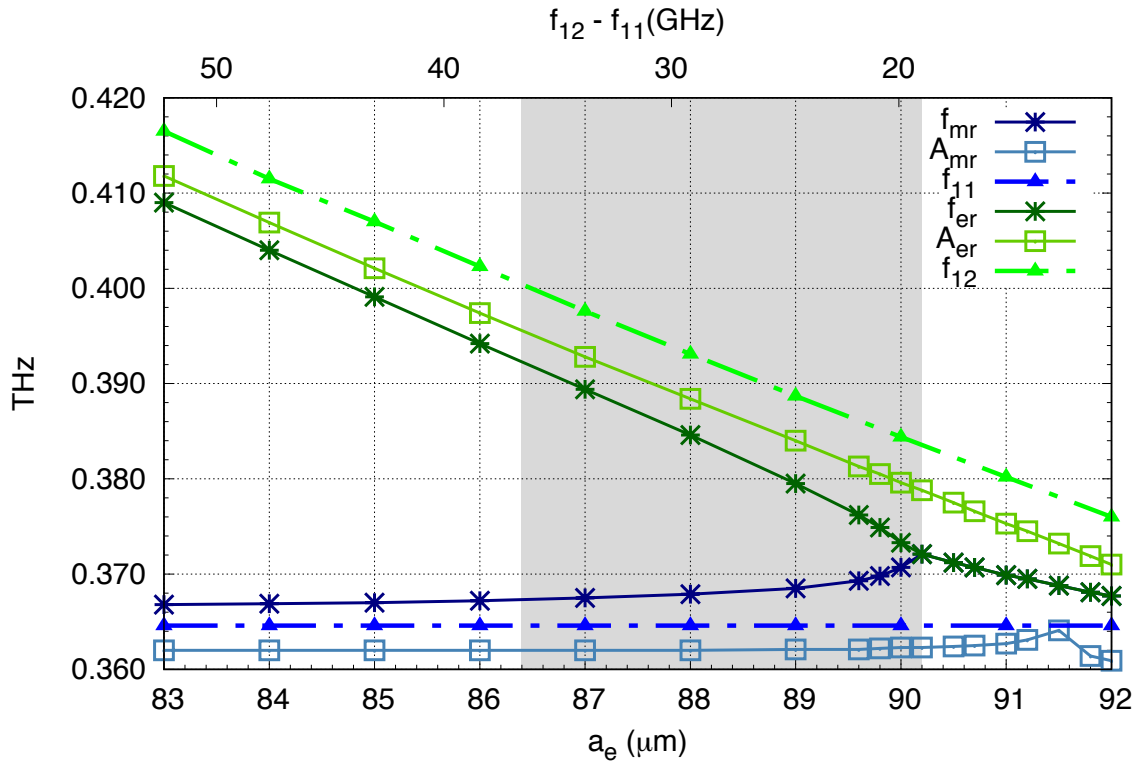}
\caption{\textit{Frequency} mode coupling: frequency of the first two modes of Mie resonances (\fmr, \fer) in function of the side length $a_e$ of the \textit{electric} resonator, namely, the frequency of the \textit{electric} mode is varying. 
The convention is the same as in Fig.\,\ref{diapason_pas}. The side length of the \textit{magnetic} resonator is  $a_m= 60\,\mu m$ and the lattice period $l_p$ is $260\,\mu m$}
\label{diapason_cote}
\end{center}
\end{figure}

Then, we studied the mode coupling inside the ADM, namely, the unit cell now consists of the two building blocks. 
The variation of the resonance frequencies (\fmr, \fer) of both modes is reported in Fig.\,\ref{diapason_pas}\, and \ref{diapason_cote} corresponding to the \textit{spatial} mode coupling and the \textit{frequency} mode coupling, respectively.  These frequencies are still given by the minima of the $S_{12}$\,parameter\,~\cite{njp14_lippens}. Decreasing the lattice period $l_p$, i.e., the distance $p_2$ between the resonators, increases the mode coupling. Varying the overlapping of the two modes stems from the decreasing of the frequency of the \textit{electric} mode, which also increases the mode coupling~\cite{acie49_giessen}. The curves are shaped as ``tuning forks'' and show that the frequencies (\fmr, \fer) of the two modes of resonance are moving closer as the mode coupling increases. To highlight this effect, the frequencies ($f_{11}$, $f_{12}$) of the resonance modes of the individual resonator are again shown in these figures.
For both the \textit{magnetic} mode and the \textit{electric} one, the distance between
the frequency (\fmr, \fer) of the resonator inside the ADM 
and 
the frequency ($f_{11}$, $f_{12}$) of the individual one respectively 
increases as the mode coupling increases. This anew evidences the mode coupling inside the ADM. 
Theses curves also evidence that further increasing the mode coupling gives rise to a frequency degeneracy  in both cases of mode coupling, that is, the two resonance frequencies of the two modes become equal, \fmr = \fer.

\begin{figure}
\begin{center}
\begin{tabular}{cc }
\includegraphics[width=0.55 \textwidth, keepaspectratio]{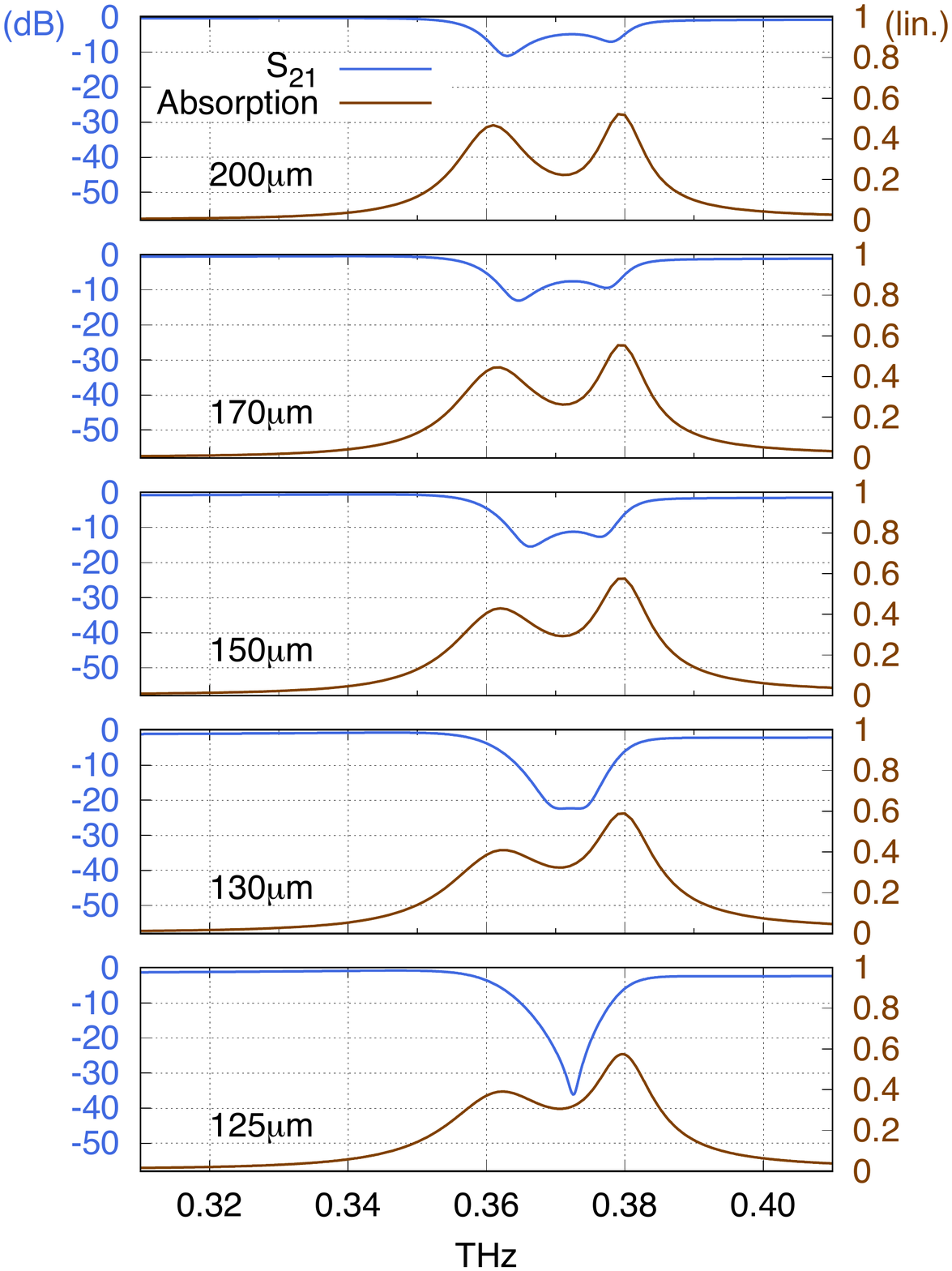} &
\includegraphics[width=0.55 \textwidth, keepaspectratio]{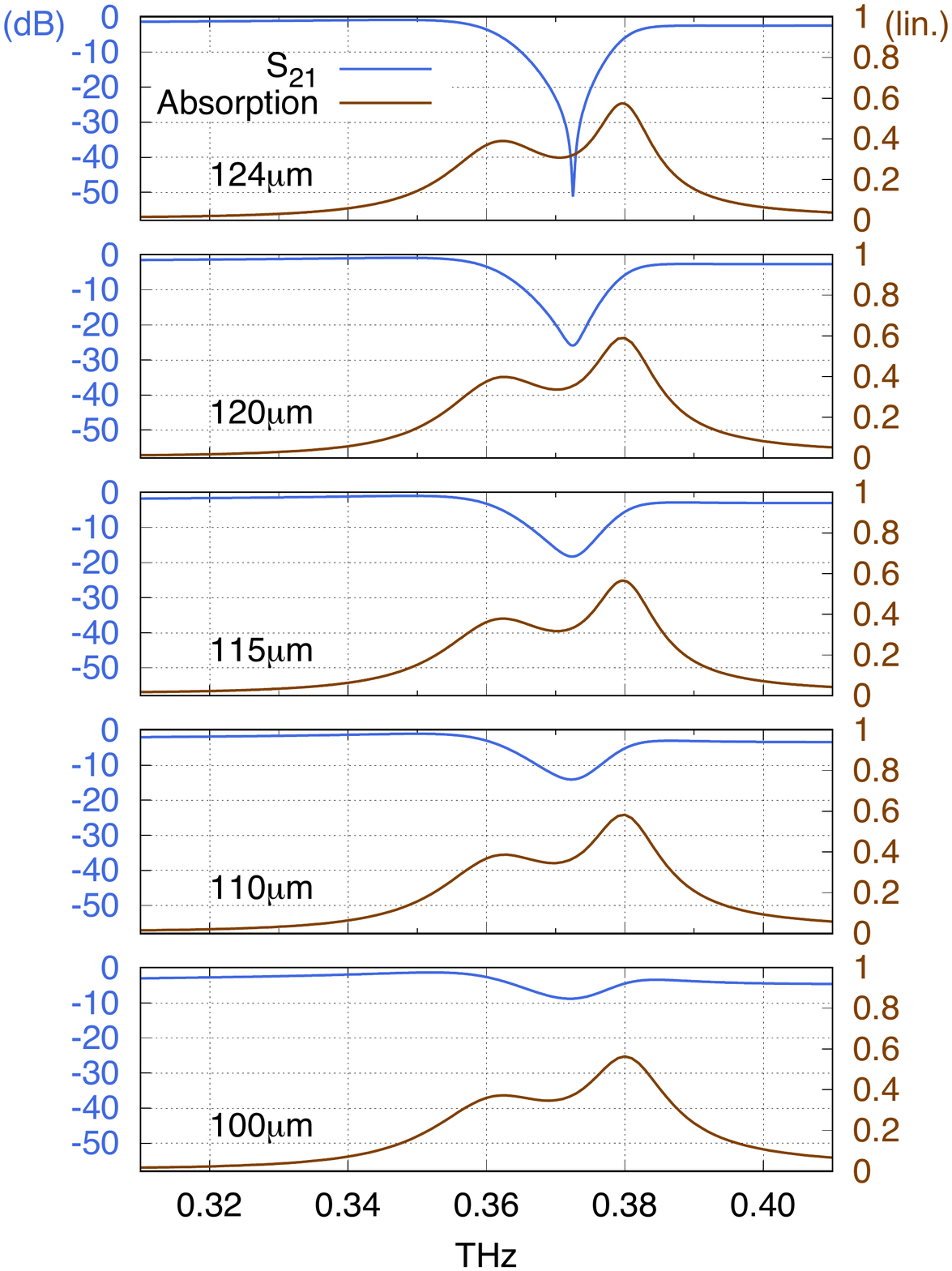}\\
(a) & (b)\\
\end{tabular}

\end{center}

\caption{Effect of the \textit{spatial} mode coupling: $S_{12}$\,parameter (log. scale (dB)) and absorption $A$ (linear scale (lin.)) for several values of the distance $p_2$ between two resonators. The side lengths of the resonators are $a_m= 60\,\mu m$ and $a_e= 90\,\mu m$.  (a) out of the frequency degeneracy regime ($125 \leq p_2 \leq 200\,\mu m$); (b) in the frequency degeneracy regime ($100 \leq p_2 \leq 124\,\mu m$). The merged minima of the $S_{12}$\,parameter attain very weak values ($\lesssim$ -51dB) when $p_2 = p_{2_{c}} = 124\,\mu m$, which correspond to the minimum of the refractive index $n_{eff_{min}}\xspace$ = -2.2.}
\label{sparam_pas}
\end{figure}

The $S_{12}$\,parameter and the absorption $A$  are reported in Fig.\,\ref{sparam_pas} in function of the frequency for several values of the lattice period $l_p$, which is relative to the \textit{spatial} mode coupling. 
Two ranges of the lattice period  $l_p$ are considered, one is out of the frequency degeneracy regime ($250 \leq l_p \leq 400\,\mu m$) and the second one in the frequency degeneracy regime ($200 \leq l_p \leq 248\,\mu m$). 
It can be noticed that the frequency of both maxima of the absorption are practically constant, whereas the frequency of both minima of the $S_{12}$\,parameter are varying according to the lattice period~$l_p$. These latter get closer as the lattice period $l_p$ increases, until to be merged, which corresponds to the frequency degeneracy. The crossing point is reached when the lattice period is equal to $l_{pc} = 250\,\mu m$. The merged minima of the $S_{12}$\,parameter attain very weak value ($\lesssim$ -51dB) for $l_p = 248\mu m$, which corresponds to the minimum of the effective index $n_{eff_{min}}(l_p = 248\mu m ) \xspace$ = -2.2. 
(See effective index curves inserted in Fig.\,2 and Fig.\,3 which are continuous, but are on the limit of continuity.)
For greater values of the mode coupling, that is, for lattice period $l_p$ smaller than 248$\,\mu m$, 
the effective parameters could not be extracted, being not continuous through all frequencies~\cite{pre70_kong}; the effective refractive  index $\neff$ is then undetermined. We observed the same frequency degeneracy behavior when studying the \textit{frequency} mode coupling (results not shown). 

The mode coupling effect we report on is different from hybridization~\cite{s302_prodan} which is observed with plasmonic metameterial~\cite{ np3_giessen,oe15_giessen}, split-ring resonators metametarials~\cite{prb83_derosny}, inductor-capacitor resonators~\cite{apl101_soukoulis}, cut wires~\cite{am19_giessen}, nanowires~\cite{nl8_christ}, nano-rings~\cite{nc3_kante}, nano-particle dimers~\cite{nl4_prodan} or silicon nanoparticles~\cite{ o3_polman}. In the latter cases, the coupling between the two identical meta-atoms which constitute the dimer leads from a trapped mode to the formation of new hybridized modes because it lifts the degeneracy of the mode of the individual meta-atoms.
Hybridization may be used to yield negative refractive index\cite{apl101_soukoulis, prb83_derosny, nc3_kante}. The unit cell is then a homo-dimer and the negative effective  index is achieved by playing with the mode coupling so as to overlap two hybridized modes which are of different kind: \textit{magnetic} or \textit{electric}. 
In the case we report on, the mechanism is different since it takes the reverse way, because 
increasing the mode coupling leads from two separated modes to a trapped one. The unit cell of our ADM is a hetero-dimer because the two meta-atoms are not identical and moreover, each one resonates in a different kind of mode: the \textit{magnetic} one and the \textit{electric} one.
Hence, increasing the mode coupling leads to the trapped mode putting together the two separated modes. Nevertheless, the mode coupling has to be strong enough to ensure negative effective  index, the unit cell being either a homo-dimer or a hetero-dimer. 

\subsection{Effective parameters}
\begin{figure}
\begin{tabular}{cc }
\includegraphics[width = 0.5\linewidth]{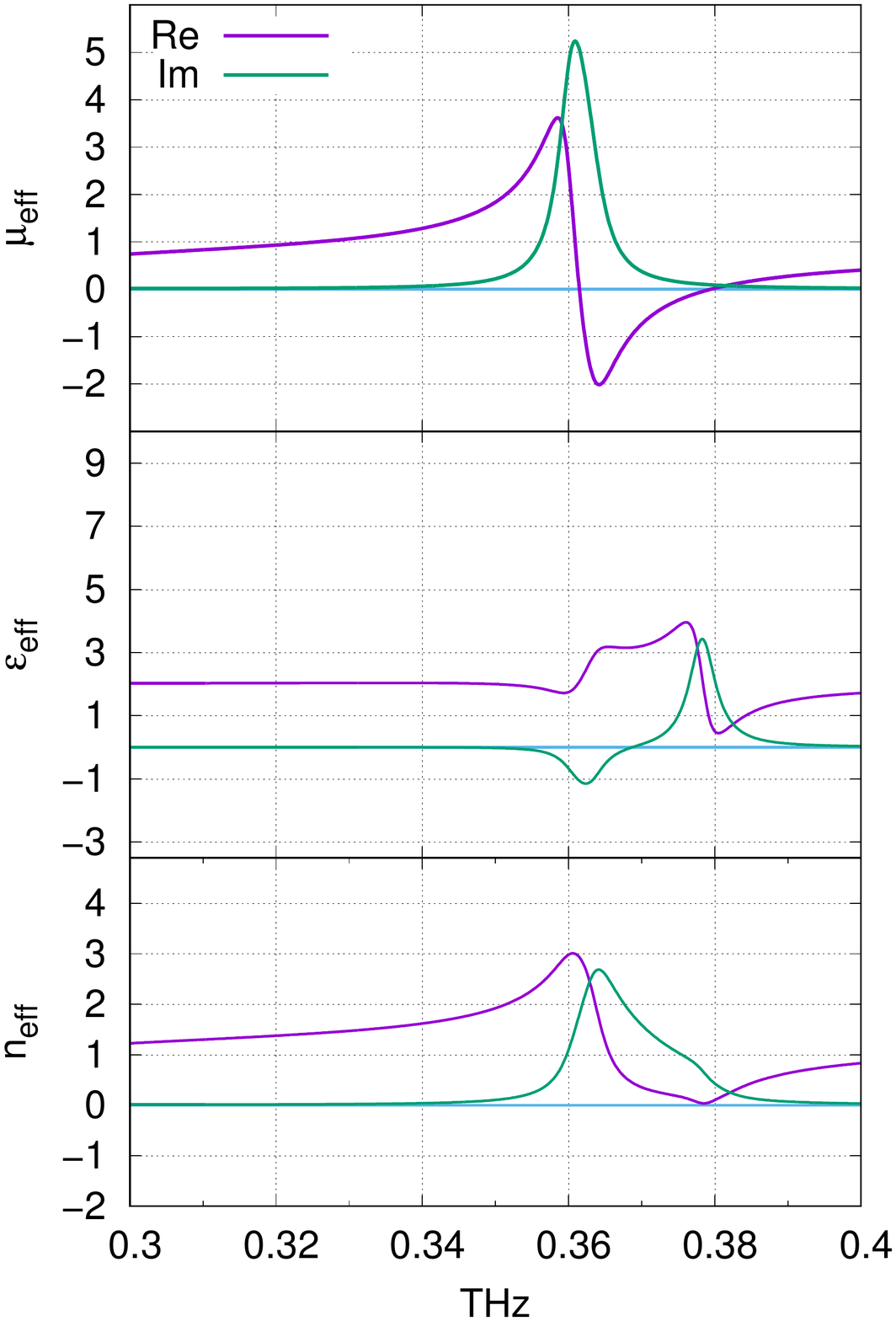}  & 
\includegraphics[width = 0.5\linewidth]{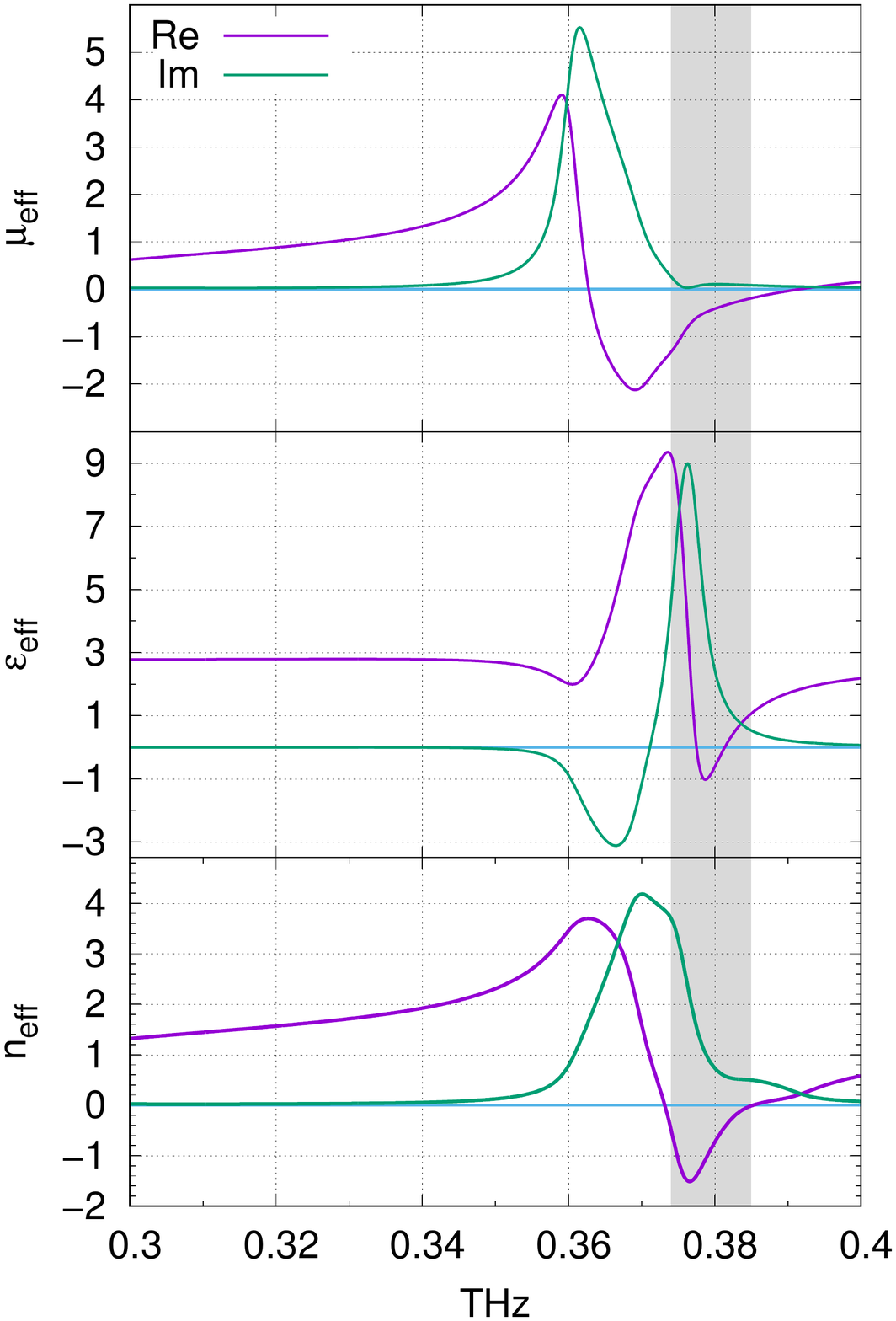}\\
(a) & (b)\\
\end{tabular}
\caption{Effect of the \textit{spatial} mode coupling: effective electromagnetic parameters (real and imaginary parts): permeability $\mu_{eff}$ (top), permittivity $\epsilon_{eff}$ (middle)  and effective index $n_{eff}$ (bottom).  (a) low mode coupling: lattice period $l_p$ = 360\,$\mu$m: the minimum value of the effective refractive  index is $n_{eff_{min}}\xspace \lesssim 0.04$ ; (b) strong mode coupling: lattice period $l_p$ = 260\,$\mu$m. The shaded area denotes  the bandwidth of negative effective index: the minimum value of the effective refractive  index is $n_{eff_{min}}\xspace$ =-1.5.}
\label{param_eff}

\end{figure}

In our ADM, a \textit{magnetic} moment ensues from the \textit{magnetic} mode giving rise to resonant effective permeability $\mu_{eff}$. Similarly, an \textit{electric} moment ensues from the \textit{electric} mode giving rise to resonant effective permittivity $\epsilon_{eff}$. The two of them are perpendicular to each other (see e.g.,\,Fig.1 in reference\,\cite{jap109_lepetit}). 
The mode coupling arises from the interaction between these electromagnetic moments and it changes with the distance between them. The side lengths of the resonators are afresh $a_m= 60\,\mu m$ and $a_e = 90\,\mu m$. 
Both resonances consequently modify the effective refractive  index $n_{eff}(\omega)$ of the ADM. The effective electromagnetic parameters $\mu_{eff}$, $\epsilon_{eff}$ and $n_{eff}$ are reported in Fig.\,\ref{param_eff} relative to two values of the lattice period $l_p = 360\,\mu$m and $l_p = 260\,\mu$m. 
They are  extracted from the $S$-parameters using the common retrieval method described in ref.\,\cite{IM19_nicolson, pi3e62_weir, mtt58_szabo, mtt55_varadan, prb65_soukoulis, pre68_soukoulis, pre70_kong}. 
The antiresonance behavior of the effective permittivity $\epsilon_{eff}$ around the \textit{magnetic} mode frequency, which is inherent in metamaterials, can be observed~\cite{pre68_soukoulis, pre71_soukoulis, prb68_alu, jo13_simovski}. 
In the former case (low mode coupling), the effective index $\neff$ does not reach negative values, whereas it does in the latter case (strong mode coupling), then satisfying equation\,\ref{nneg}. The minimum value of the effective index is then $n_{eff_{min}}(l_p = 260\,\mu m)= -1.5$. In the former case, the effective index $\neff$ is below unity and close to zero. Its minimal value is $n_{eff_{min}}(l_p = 360\,\mu m) \lesssim  0.04$, evidencing that adjusting the mode coupling makes ADMs suitable for epsilon-near-zero (ENZ) metamaterials~\cite{pre70_ziolkowski, prb75_alu} (see also Fig.\,\ref{n_min_spatial}), or even null effective index~\cite{np7_valentine}.
It can also be noticed that the mode coupling strongly enhances the amplitude of both resonances, notably the \textit{electric} one, and that it brings closer the two modes.
Besides, as we are dealing with a high refractive index bulk material (\ntio$\simeq$ 10), the wavelength inside it is about one-tenth of that in vacuum. Consequently, we calculated the static effective permittivity, i.e., beyond the resonances, from the Maxwell-Garnett model\,~\cite{pier51_sihvola}. It is equal to $\epsilon_{eff} = 2.1$ and $\epsilon_{eff} = 2.9$  corresponding to both values of the lattice period $l_p = 360\,\mu$m and $l_p = 260\,\mu$m, respectively, which is in good agreement with the results of the simulation. 

To engineer the electromagnetic properties of an ADM, one can consequently play with either mode couplings:  the \textit{spatial} mode coupling or the \textit{frequency} mode coupling. Fig.\,\ref{coupling} gather the role of both mode couplings: the stronger the two mode couplings, the more negative the effective index and the larger the bandwidth of negative effective  index (cf. Fig. \ref{n_min_spatial} and \ref{n_min_freq}). Combining the two mode couplings, negative effective  index as low as $\neff=-2.8$ is obtained.

\subsection{Dielectric function, phonons and strontium titanate (\sto)}
Others high permittivity materials, having low losses, can be investigated to study the mode coupling inside ADMs at terahertz frequencies, e.g., \sto\,\cite{ pr126_spitzer, pr145_barker, apl90_han, jjap48_matsumoto}.
However, the dielectric function $\epsilon_r(\omega)$ of these high permittivity materials is dispersive,  because of the lattice vibrations, namely, the optical phonons~~\cite{prb10_gervais}. Their frequency is in the THz range, and we are concerned by the transverse optical phonon of lowest frequency (TO$_1$). The TO$_1$ phonon frequency of \sto{} is 2.70\,THz\cite{pr126_spitzer}, while that of \tio{} is 5.6\,THz\,\cite{pr126_spitzer, prb10_gervais}. 
The dielectric function $\epsilon_r(\omega)$ is described by the classical oscillator model or the Four-Parameter Semi Quantum (FPSQ) model\cite{pr126_spitzer, prb10_gervais}. Measurements at THz frequency, reported in the literature, are in good agreement with these models for {\tio\cite{jjap47_matsumoto, mtt53_berdel} and \sto\cite{jjap48_matsumoto, apl91_tsurumi, apl90_han, prb64_petzelt}. The two yield the same dielectric function $\epsilon_r(\omega)$ at the operating frequency. 
We also simulated a similar ADM but consisting of HPRs made of \sto{} and found that it exhibits the same mode coupling effect (results not shown herein). The effective index reaches values as low as $n_{eff_{min}} = -7$, meaning that the effect of the mode coupling is stronger when a higher permittivity material is used ($\epsilon_r \simeq 300$~\cite{jjap48_matsumoto} instead of $\epsilon_r = 94$), because this heightens the resonances. 

However, losses (imaginary part of $\epsilon_r(\omega)$) resulting from the TO$_1$ phonon greatly increase and therefore limits the operating range at terahertz frequencies. Besides, the real part of the dielectric function $\epsilon_r(\omega)$ falls down at higher frequency~\cite{apl91_tsurumi, jjap48_matsumoto}, which drastically modifies the Mie resonances. The lower permittivity of \tio{} leads to greater side lengths $a_m$ and $a_e$ of each resonator (see equation\,\ref{eq2}) and therefore, it facilitates their fabrication. 
Consequently, \tio{} is more suitable for ADM applications at THz frequencies.  

\begin{figure}
  \includegraphics[width= \linewidth, trim = 0cm 0cm 0cm 0cm, clip = true]{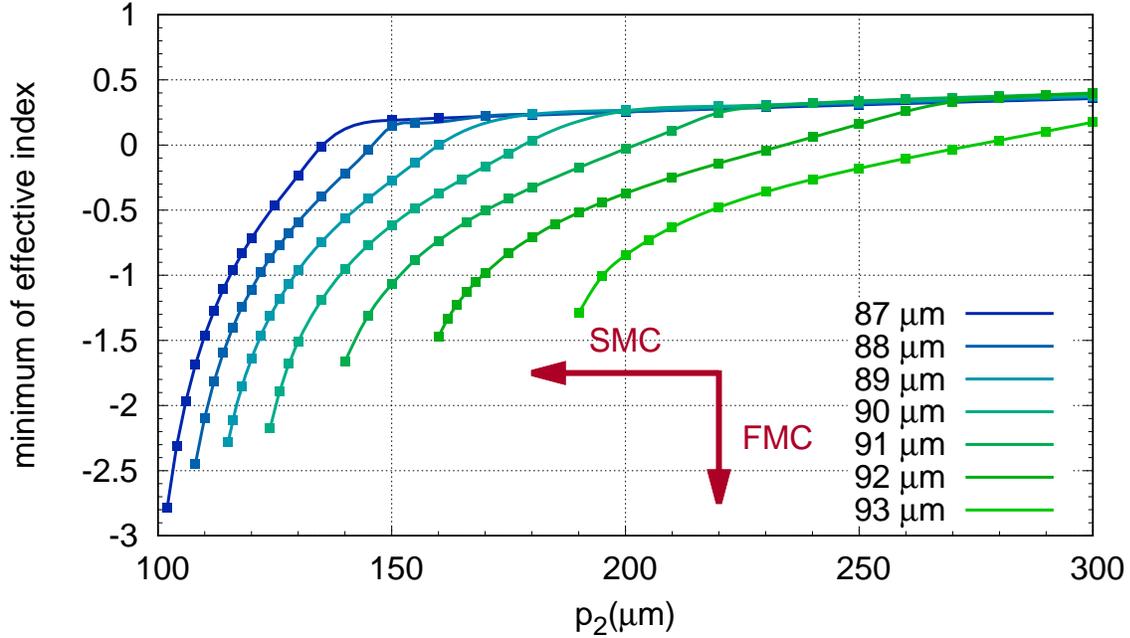}
\caption{Effect of both mode couplings (\textit{spatial} mode coupling (SMC) and frequency mode coupling (FMC)): minimal value of the effective refractive index $n_{eff}$ in function of the distance $p_2$ between two resonators for several values of the side length $a_e$ of the \textit{electric} resonator. The side length of the \textit{magnetic} resonator is $a_m= 60\,\mu m$. The two arrows denote increasing mode coupling. The minimum of the effective index is $n_{eff_{min}}\xspace$ = -2.8. }
\label{coupling}
\end{figure}

\section{Conclusion}
We have studied mode coupling effects in ADMs at terahertz frequencies and we show that the mode coupling has to be sufficiently strong to ensure negative effective  index of refraction. Tuning the first two modes of Mie resonances of an ADM by adjusting the mode coupling allows to set the effective index from a near-zero value to a negative value. 
We studied both spatial mode coupling and frequency mode coupling. Increasing both brings the modes closer until they are merged. Thus, we highlight the frequency degeneracy of the first two resonance modes, namely, the two frequencies are equal, and the effective index is then undetermined. At the crossing point, the effective index reaches its lowest value.

\section{Author contributions}
\'E.A. supervised the study, analyzed the results and wrote the paper. S.M. made the simulation and  analyzed the results.

\ack

The authors thank Aloyse Degiron for helpful discussion. \'E.A.  thanks Fabrice Rossignol  and Jean-Pierre Ganne for their help. This work has been funded by the Agence Nationale de la Recherche \textit{T\'eraM\'etaDiel} project (number\,ANR--12-BS03-0009)

\newpage
\bibliographystyle{iopart-num.bst}
\bibliography{adm_coupling_arxiv.bib}

\end{document}